\documentclass[preprint,showpacs,preprintnumbers,amsmath,amssymb]{revtex4}

\usepackage{graphicx}% Include figure files
\usepackage{dcolumn}% Align table columns on decimal point
\usepackage{bm}% bold math
\usepackage{longtable}
\usepackage{epsfig}
\def\be{\begin{equation}}
\def\ee{\end{equation}}

\def\gtrsim{\,\,\raise0.14em\hbox{$>$}\kern-0.76em\lower0.28em\hbox
{$\sim$}\,\,}
\def\lesssim{\,\,\raise0.14em\hbox{$<$}\kern-0.76em\lower0.28em\hbox
{$\sim$}\,\,}

\begin{document}
%\draft
%
\title{The reaction $^{13}$C($\alpha$,n)$^{16}$O: a
background for the observation of geo-neutrinos}
\author{S. Harissopulos$^{1}$, H. W. Becker$^{2}$, J. W. Hammer$^{3}$,
        A. Lagoyannis$^{1}$, C. Rolfs$^{4}$,  F. Strieder$^{4}$}
\affiliation{$^{1}$Institute of Nuclear Physics, NCSR``Demokritos",
            153.10 Aghia Paraskevi, Athens, Greece}
\affiliation{$^{2}$Dynamitron-Tandem-Laboratorium, Ruhr-Universit\"at Bochum,
%             Universit\"atsstrasse 150,
             44801 Bochum, Germany}
\affiliation{$^{3}$Institut f\"ur Strahlenphysik, Universit\"at Stuttgart,
%             Allmandring 3,
             70569 Stuttgart, Germany}
\affiliation{$^{4}$Institut f\"ur Physik mit Ionenstrahlen,
             Ruhr-Universit\"at Bochum,
%             Universit\"atsstrasse 150,
             44801 Bochum, Germany}
\date{\today}
%
%\vspace*{-0.5cm}
\begin{abstract}
The absolute cross section of the $^{13}$C($\alpha$,n)$^{16}$O reaction
has been measured at E$_{\alpha}$ = 0.8 to 8.0 MeV with an overall
accuracy of 4\%. The precision is needed to subtract reliably a
background in the observation of geo-neutrinos, e.g. in the KamLAND
detector.
\end{abstract}
\maketitle
% body of paper here
%\vspace*{-0.5cm}
\section{Introduction}
\label{Sect:1}
In order to understand better the energy source of the earth, e.g.
the radioactive decay of U, Th and others, the KamLAND neutrino
detector is planning to observe the geo-neutrinos from these decays.
The antineutrinos $\nu_{a}$ from the decays in the earth interact
with the protons in the liquid scintillator of KamLAND via the
inverse $\beta$ decay $\nu_{a}$+p$\rightarrow$n+e$^{+}$ and are
detected with phototubes via the combination of (i) a prompt signal
from the slowing down and annihilation of the e$^{+}$ and (ii) a
delayed signal provided by the 2 MeV $\gamma$-ray from the
p(n,$\gamma$)d reaction, which occurs after the neutron has
been thermalized.

The KamLAND detector consists of 1 kilotons of organic
scintillator and contains thus about 10 tons of $^{13}$C. The
detector contains also 10$^{-20}$ g/g of $^{210}$Pb nuclides
(halflife = 22.3 y), which decay via $^{210}$Bi and $^{210}$Po
into $^{206}$Pb emitting in the process a 5.3 MeV $\alpha$-particle.
When these $\alpha$-particles slow down in the scintillator, they
can initiate the $^{13}$C($\alpha$,n)$^{16}$O reaction (Q = 2.215 MeV),
whereby the fast neutrons perfectly mimic the antineutrino signal.
Firstly, the few MeV neutron transfers its kinetic energy to the
protons in the medium, such that a few MeV proton emerges from this
first collision. The proton in turn produces ionisation, which
results in a prompt signal simulating (i). A prompt signal (i)
can also be created in the slowing down (ionization) of the 5.3 MeV
alpha-particle. Secondly, as the neutron
is slowed down it will create a signal like (ii). In this way a fake
antineutrino signal is built.

The $^{13}$C($\alpha$,n)$^{16}$O reaction creates an important
background and gives a number of fake events which are comparable
to those expected from geo-neutrinos \cite{1}. For detecting geo-neutrinos
one needs to subtract the fake events in a reliable way. With the
present 30\% uncertainty in the absolute cross section $\sigma$ of
$^{13}$C($\alpha$,n)$^{16}$O over the relevant energy range
(measured in \cite{2} at E$_{\alpha}$ = 1.0 to 5.4 MeV), it appears
impossible \cite{1} to extract a geo-neutrino signal from any detector
at any place in the world.
Calculations indicate \cite{1} that for a 10\% uncertainty in $\sigma$
the discovery of geo-neutrinos will be achievable.
In recent years new precise $\sigma$ measurements have been carried out
at low energies \cite{3,4}, i.e. below E$_{\alpha}$ = 1.1 MeV.
We report on new precise $\sigma$ measurements in the energy range
E$_{\alpha}$ = 0.8 to 8.0 MeV.
\section{Equipment and procedures}
\label{sec:2}
The tandem accelerator of the Dynamitron - Tandem - Laboratorium (DTL)
at the Ruhr-Universit\"at Bochum provided the $^{4}$He beam, with
a current on target of less than 100 nA. The absolute energy of the
beam was checked to a precision of $\pm$3 keV; the energy spread was
0.8 keV at E$_{\alpha}$ = 1.0 MeV. The beam passed through
collimators of \O~ = 15, 3, and 5 mm diameter at a distance from
the target of 3.06, 1.43, and 0.80 m, respectively, and was stopped
at the target. The air-cooled target (\O~ = 40 mm) was installed
in an electrically insulated pipe of 0.80 m length (\O~ = 40 mm),
which served as a long Faraday cup. A suppression voltage of 300 V
was applied to the last aperture in front of the target pipe and
this led to a precision of 2\% in the charge measurement at the
target. The vacuum was better than 3$\times$10$^{-7}$ mbar.

The 22 $\mu$g/cm$^{2}$ thick C target (on a 0.2 mm thick Ta backing)
was enriched to 99\% in $^{13}$C. The enrichment was confirmed within
$\pm$2\% using the narrow resonance at E$_{pr}$ = 1.748 MeV in
$^{13}$C(p,$\gamma$)$^{14}$N ($\gamma$-ray detection with a 12$\times$12
inch NaI crystal in close geometry, placed at another beam line) and
comparing the thick-target $\gamma$-ray yields obtained for the above
enriched target as well as for a C target of natural isotopic composition
($^{13}$C = 1.1\%). The results are shown in Fig. 1, where the enriched
target was analysed at the completion of the $^{13}$C($\alpha$,n)$^{16}$O
experiment: the 50\% yield-point is identical for both targets indicating
no detectable deposition of natural carbon on the enriched target. The
deduced target thickness for the enriched target (Fig. 1) was
$\Delta_{p}$ = 3.25$\pm$0.10 keV, consistent with the value
$\Delta_{p}$ = 3.14$\pm$0.10 keV obtained initially. The results for the
E$_{\alpha r}$ = 1.054 and 1.336 MeV narrow resonances in
$^{13}$C($\alpha$,n)$^{16}$O are shown in Fig. 2 leading to
$\Delta_{\alpha}$ = 37.0$\pm$1.0 and 33.2$\pm$1.0 keV,
respectively. Using stopping power tables \cite{5} (with a 3\%
uncertainty \cite{4}) one arrives at an average value for the $^{13}$C
areal target density of N$_{{\rm t}}$ = (10.21$\pm$0.31)$\times$10$^{17}$
atoms/cm$^{2}$, where the quoted error arrives from the uncertainty
in the stopping power.

The neutron yields were measured using a 4$\pi$ detector consisting
of thermal-neutron detectors, i.e. 16 $^{3}$He-filled proportional
counters (sensitive length = 45.7 cm, \O~ = 2.54 cm, pressure = 4 bar),
embedded in a polyethylene-cylinder moderator (length = 1.00 m,
\O~ = 35.0 cm) with a central bore hole (\O~ = 7.0 cm) to install the
target. At an inner circle of \O~ = 16.0 cm there were 8 $^{3}$He
counters and at \O~ = 24.0 cm there were another 8 $^{3}$He counters,
all embedded in holes of \O~ = 3.0 cm in the moderator. The inner and
outer counters are positioned with a relative angle difference of
22.5$^{\circ}$.
To reduce the neutron background arising from cosmic showers, several
components for passive shielding (Cd, polyethylene, bor-polyethylene,
bor-paraffin) are placed around the moderator leading to a background
rate of 0.22 counts/s. In the present experiment the spectra of the
16 counters were summed and stored in an ADC. To monitor dead time
effects (kept below 4\%), a pulser was used.

The neutron efficiency $\eta_{n}$(E$_{n}$) as a function of neutron
energy E$_{n}$ (in MeV) was calculated with the Monte Carlo program
MCNP leading to the expression \cite{6}
\begin{equation}
\eta_{n}({\rm E}_{n}) = 0.7904+0.1607({\rm E}_{n}+247.07)
                  exp(-0.114{\rm E}_{n}) [\%]
\end{equation}
at E$_{n}$ = 2.0 to 9.0 MeV. With a calibrated ($\pm$1.1\%)
$^{252}$Cf source, the neutron detection efficiency at the mean energy
E$_{n}$ = 2.3 MeV was found \cite{6} to be 32.1$\pm$0.5\% in good
agreement with the calculated value 31.6\%. We used the calculated
curve to correct the observed neutron yields for efficiency: over the
energy range of the present experiment, E$_{\alpha}$ = 0.8 to 8.0 MeV,
$\eta_{n}$(E$_{n}$) varies from 31\% to 16\% for a mean
neutron emission angle of 90$^{\circ}$.

\section{Results}
\label{Sect:3}
An excitation function was obtained with the $^{13}$C target at
energies between E$_{{\rm \alpha}}$ = 0.8 and 8.0 MeV in steps
of $\delta$ = 10 keV, where the target thickness $\Delta$(E$_{{\rm \alpha}}$)
varied between 39 and 13 keV, respectively. The exceptions in steps
are for E$_{{\rm \alpha}}>$7.2 MeV ($\delta$ = 20 keV) and near
narrow resonances ($\delta$ = 0.2 keV). The statistical error of
each data point was better than 0.2\%.  After completion
of the excitation function, yield tests were performed at selected
energies over the total energy range leading to a reproducibility
of better than 2\% (=relative error).

For non-resonant energies, where the yields change slowly with energy,
the observed yields (and thus the cross sections) are associated with
an effective energy E$_{{\rm eff}}$ over the target thickness
$\Delta$(E$_{{\rm \alpha}}$) \cite{7}. For energies with resonance
structures, where the resonance width is comparable to the target
thickness, the deduced value E$_{{\rm eff}}$ and associated yield
represents a mean cross section over $\Delta$(E$_{{\rm \alpha}}$).

The observed neutron yield N$_{{\rm n}}$ at a given E$_{{\rm eff}}$
is related to the cross section $\sigma$(E$_{{\rm eff}}$) via the
relation \cite{7}
\begin{equation}
N_{\rm n} = N_{{\rm \alpha}} N_{{\rm t}} \eta_{{\rm n}}
              \sigma(E_{{\rm eff}})
\end{equation}
where N$_{{\rm \alpha}}$ is the number of incident ${\rm \alpha}$
projectiles. At E$_{{\rm eff}}$ = 1.00 MeV we found
$\sigma$(E$_{{\rm eff}}$) = 157$\pm$7 $\mu$b, where the error
corresponds to the quadratic sum of the uncertainties in
N$_{{\rm n}}$ (1.6\%), N$_{{\rm \alpha}}$ (2.0\%),
N$_{{\rm t}}$ (3.0\%) and $\eta_{{\rm n}}$ (1.7\%). Our value is in
good agreement with the previous value of 146$\pm$7 $\mu$b \cite{4}
leading to a weighted average of 152$\pm$5 $\mu$b, which we adopted
as reference value.% (Table 1 and Fig. 3).
With the relative error of 2\% and this absolute error of 3\%
one arrives at an overall accuracy of 4\% for the data of the present
work.

For the case of the narrow resonances at E$_{{\rm \alpha}r}$ = 1.054
and 1.336 MeV (Fig. 2) and 1.590 MeV (not shown), the observed neutron
yields lead to resonance strengths $\omega \gamma$, which represent the
integral over the resonance width $\Gamma$: the respective widths are
$\Gamma$ = 1.5, 0.6, and 1.0 keV \cite{2} and the respective strengths
are $\omega \gamma$ = 12.1$\pm$0.6, 33.3$\pm$1.8, and 11.5$\pm$1.2 eV
from the present work and $\omega \gamma$ = 11.9$\pm$0.6, not given, and
10.8$\pm$0.5 eV from \cite{4} leading to the accepted values of
$\omega \gamma$ = 12.0$\pm$0.4, 33.3$\pm$1.8, and 10.9$\pm$0.5 eV.
With these quantities we calculated the cross section
$\sigma$(E$_{{\rm \alpha}r}$) at the energy region of these resonances.

Our cross-section results plotted in Fig. 3 are also given in Table I.
The present absolute cross sections for the non-resonant data are
lower by about 30\% compared to the values reported in previous work
\cite{2}: from the data shown in Fig. 3 of \cite{2} we find
$\sigma_{present}$/$\sigma_{[2]}~\cong~$0.63, 0.68, 0.64, and 0.72 at
the broader resonances E$_{\alpha}$=2.3, 3.3, 4.4, and 5.0 MeV, respectively.
\section{Conclusions}
\label{Sect:4}
With an overall accuracy of 4\% for the absolute cross section of
$^{13}$C($\alpha$,n)$^{16}$O one can subtract reliably
this background in the KamLAND detector and thus the discovery of
geo-neutrinos appears achievable.
\begin{acknowledgments}
This work supported in part by the German Academic Exchange Service
(DAAD), the Hellenic State Scholarships Foundation (IKY), and the
Dynamitron-Tandem-Laboratorium (DTL). The authors would like to
thank G. Fiorentini for suggesting the experiment and
J. Wrachtrup for the loan of the neutron detector.
\end{acknowledgments}

\newpage
\begin{figure}[h]
%\resizebox{0.35\textwidth}{!}
%  \includegraphics{Harissopulos_C13an_Fig1ps_revised.ps,width=0.5\hsize}
  \epsfig{figure=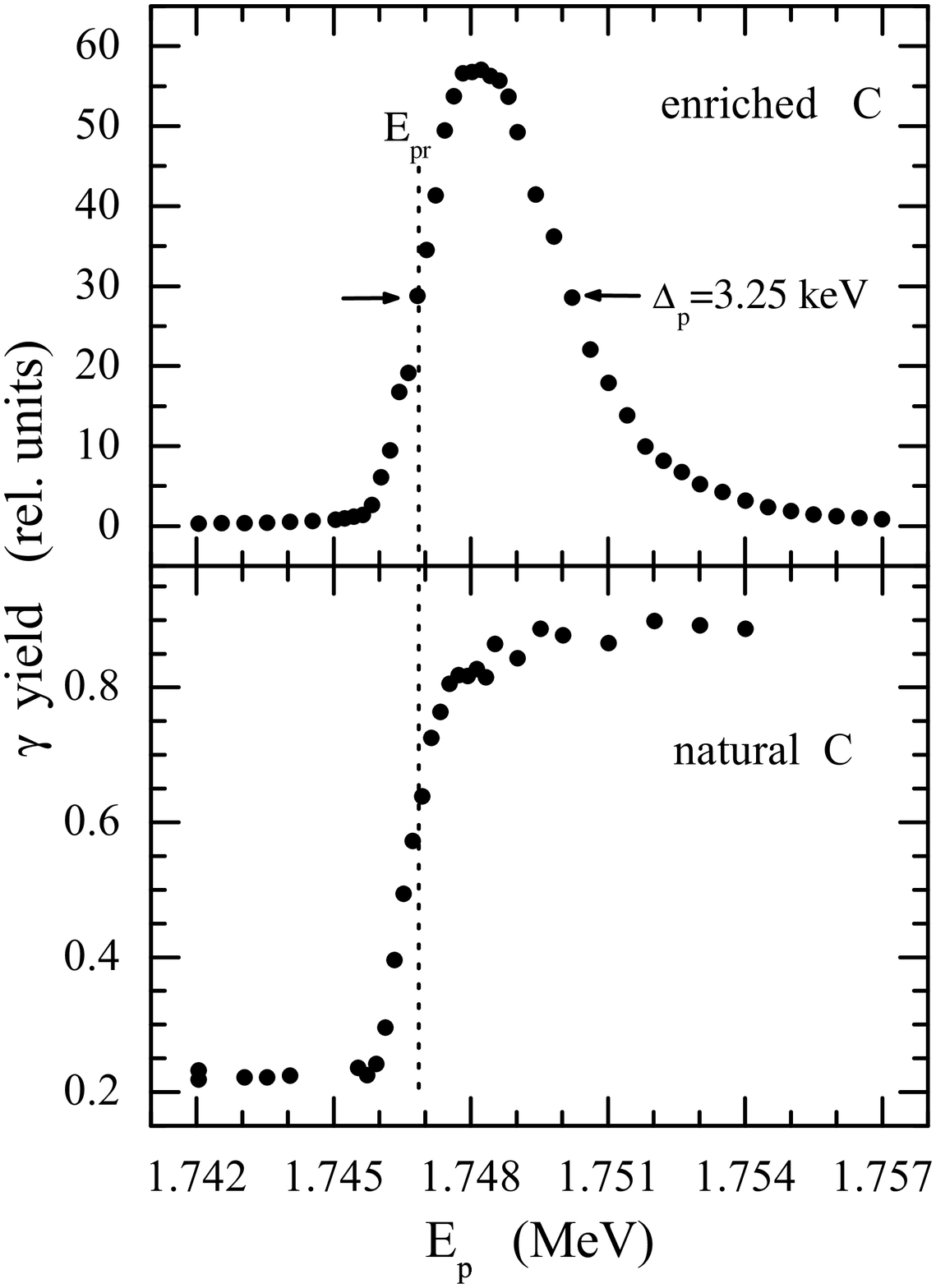,width=0.5\hsize}
%}
\label{fig1}
\caption{Thick target $\gamma$-ray yield for the narrow E$_{{\rm p}r}$ = 1.747
MeV resonance in $^{13}$C(p,$\gamma$)$^{14}$N obtained with
a C target enriched in $^{13}$C and with a C target of natural isotopic
composition.}
\end{figure}
\newpage
\begin{figure}[h]
%\resizebox{0.35\textwidth}{!}{%
%  \includegraphics{Figure2eps.eps}
  \epsfig{figure=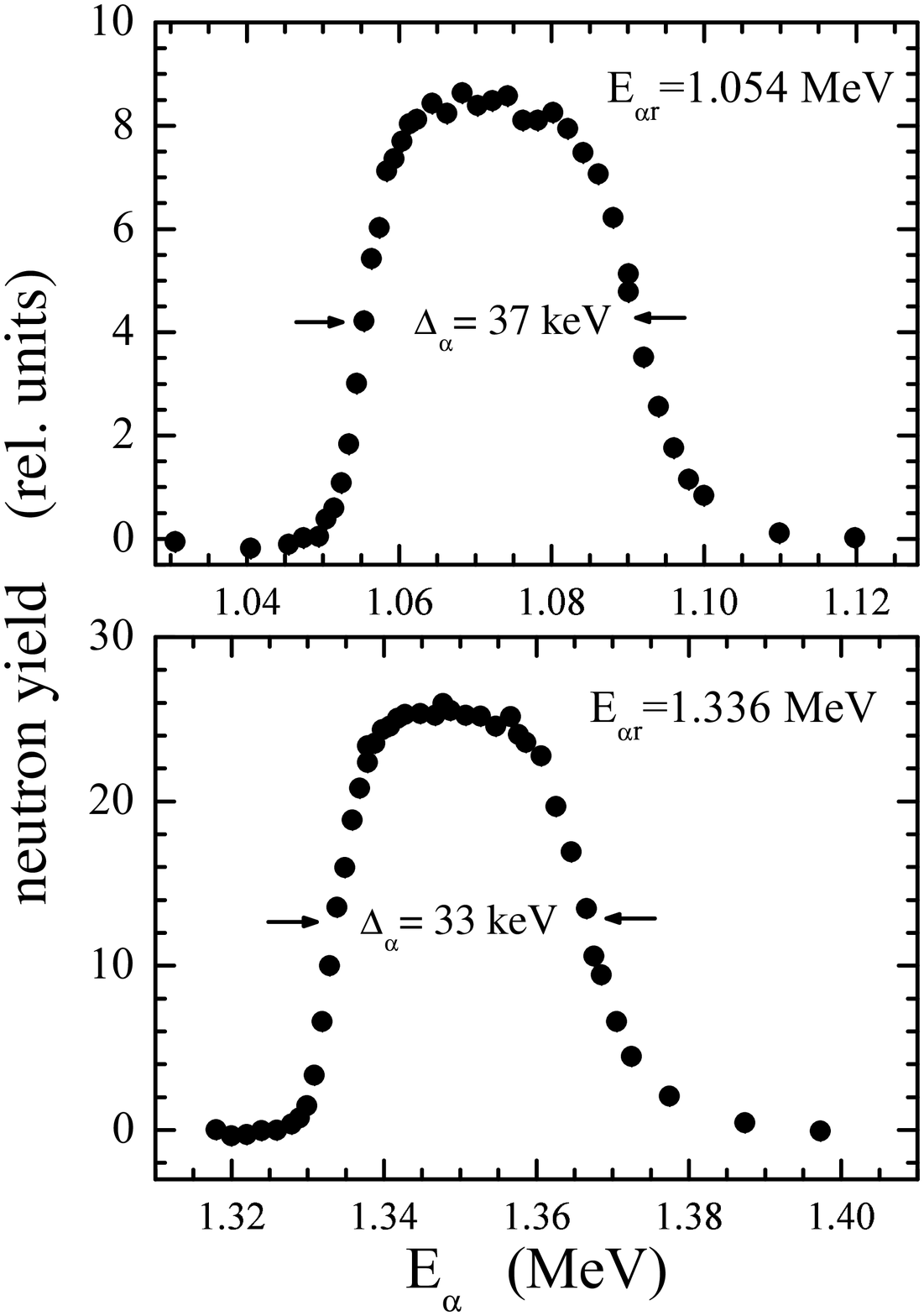,width=0.5\hsize}
%}
\label{fig2}
\caption{Thick target neutron yield for the narrow E$_{{\rm \alpha}}$ = 1.054
and 1.336 MeV resonances in $^{13}$C($\alpha$,n)$^{16}$O obtained
with a C target 99\% enriched in $^{13}$C; and non-resonant yield has been
subtracted from the data.}
\end{figure}
\newpage
\begin{figure}[h]
%\resizebox{0.75\textwidth}{!}{%
%  \includegraphics{Figure3eps.eps}
  \epsfig{figure=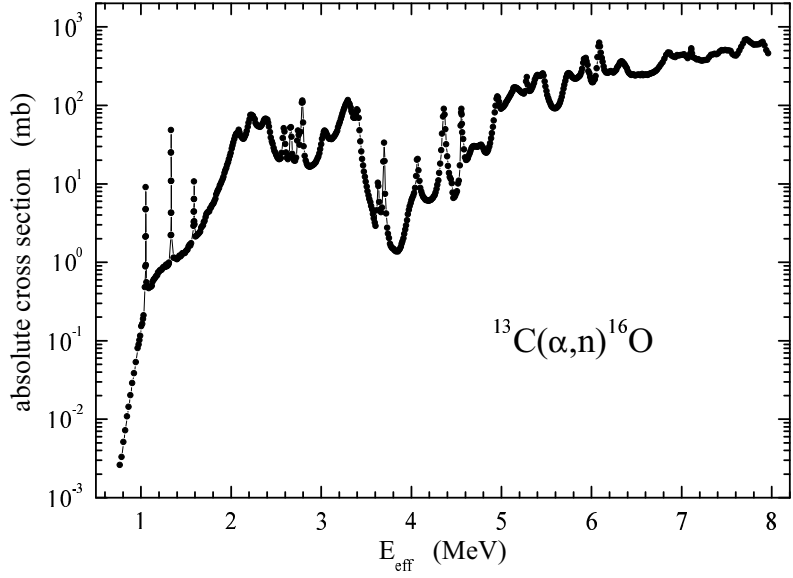,width=0.5\hsize,angle=270}
%}
\label{fig3}
\caption{Absolute cross section of $^{13}$C($\alpha$,n)$^{16}$O
between E$_{{\rm eff}}$ = 0.77 and 7.96 MeV. The solid curve is to
guide the eye only. For the narrow resonances at E$_{{\rm \alpha}r}$ = 1.054,
1.336, and 1.590 MeV the corresponding resonant cross section was deduced
from the measured strength and total width.}
\end{figure}
\newpage
\begin{center}
\begin{longtable*}{cccccccccccccccc}
\caption{Absolute cross section of $^{13}$C($\alpha$,n)$^{16}$O}
\label{tabel1} \\
\hline \hline
\multicolumn{1}{c}{E$_{\rm eff}$} &
\multicolumn{1}{c}{$\sigma$}    &
\multicolumn{1}{c}{E$_{\rm eff}$} &
\multicolumn{1}{c}{$\sigma$}     &
\multicolumn{1}{c}{E$_{\rm eff}$} &
\multicolumn{1}{c}{$\sigma$}     &
\multicolumn{1}{c}{E$_{\rm eff}$} &
\multicolumn{1}{c}{$\sigma$}     &
\multicolumn{1}{c}{E$_{\rm eff}$} &
\multicolumn{1}{c}{$\sigma$}     &
\multicolumn{1}{c}{E$_{\rm eff}$} &
\multicolumn{1}{c}{$\sigma$}     &
\multicolumn{1}{c}{E$_{\rm eff}$} &
\multicolumn{1}{c}{$\sigma$}     &
\multicolumn{1}{c}{E$_{\rm eff}$} &
\multicolumn{1}{c}{$\sigma$}    \\
\multicolumn{1}{c}{(MeV)} & \multicolumn{1}{c}{(mb)} &
\multicolumn{1}{c}{(MeV)} & \multicolumn{1}{c}{(mb)} &
\multicolumn{1}{c}{(MeV)} & \multicolumn{1}{c}{(mb)} &
\multicolumn{1}{c}{(MeV)} & \multicolumn{1}{c}{(mb)} &
\multicolumn{1}{c}{(MeV)} & \multicolumn{1}{c}{(mb)} &
\multicolumn{1}{c}{(MeV)} & \multicolumn{1}{c}{(mb)} &
\multicolumn{1}{c}{(MeV)} & \multicolumn{1}{c}{(mb)} &
\multicolumn{1}{c}{(MeV)} & \multicolumn{1}{c}{(mb)} \\ \hline
\endfirsthead
\multicolumn{16}{c}{ \tablename ~~\thetable{}: -- continued from previous page} \\ \hline \hline
\multicolumn{1}{c}{E$_{\rm eff}$} &
\multicolumn{1}{c}{$\sigma$}    &
\multicolumn{1}{c}{E$_{\rm eff}$} &
\multicolumn{1}{c}{$\sigma$}     &
\multicolumn{1}{c}{E$_{\rm eff}$} &
\multicolumn{1}{c}{$\sigma$}     &
\multicolumn{1}{c}{E$_{\rm eff}$} &
\multicolumn{1}{c}{$\sigma$}     &
\multicolumn{1}{c}{E$_{\rm eff}$} &
\multicolumn{1}{c}{$\sigma$}     &
\multicolumn{1}{c}{E$_{\rm eff}$} &
\multicolumn{1}{c}{$\sigma$}     &
\multicolumn{1}{c}{E$_{\rm eff}$} &
\multicolumn{1}{c}{$\sigma$}     &
\multicolumn{1}{c}{E$_{\rm eff}$} &
\multicolumn{1}{c}{$\sigma$}    \\
\multicolumn{1}{c}{(MeV)} & \multicolumn{1}{c}{(mb)} &
\multicolumn{1}{c}{(MeV)} & \multicolumn{1}{c}{(mb)} &
\multicolumn{1}{c}{(MeV)} & \multicolumn{1}{c}{(mb)} &
\multicolumn{1}{c}{(MeV)} & \multicolumn{1}{c}{(mb)} &
\multicolumn{1}{c}{(MeV)} & \multicolumn{1}{c}{(mb)} &
\multicolumn{1}{c}{(MeV)} & \multicolumn{1}{c}{(mb)} &
\multicolumn{1}{c}{(MeV)} & \multicolumn{1}{c}{(mb)} &
\multicolumn{1}{c}{(MeV)} & \multicolumn{1}{c}{(mb)} \\ \hline
\endhead
\hline \multicolumn{16}{c}{Continued on next page} \\ \hline
\endfoot
\hline \hline
\endlastfoot
0.767   &       0.0026  &       1.588   &       4.30    &       2.385   &       67.5    &       3.220   &       67.0    &       4.065   &       20.2    &       4.878   &       32.9    &       5.721   &       227     &       6.553   &       244     \\
0.786   &       0.0033  &       1.589   &       6.25    &       2.395   &       66.1    &       3.230   &       72.1    &       4.074   &       20.7    &       4.888   &       37.8    &       5.731   &       247     &       6.563   &       250     \\
0.807   &       0.0051  &       1.590   &       10.5    &       2.405   &       65.3    &       3.239   &       78.2    &       4.083   &       14.8    &       4.897   &       44.2    &       5.740   &       254     &       6.573   &       241     \\
0.826   &       0.0072  &       1.591   &       6.26    &       2.414   &       59.8    &       3.249   &       85.8    &       4.092   &       10.9    &       4.907   &       53.0    &       5.750   &       256     &       6.583   &       243     \\
0.846   &       0.0108  &       1.592   &       4.33    &       2.424   &       53.3    &       3.259   &       91.9    &       4.102   &       8.85    &       4.917   &       64.0    &       5.760   &       252     &       6.593   &       245     \\
0.866   &   0.0143  &   1.593   &   3.32    &   2.434   &   45.6    &   3.269   &   99.0    &   4.113   &   7.96    &   4.927   &   80.7    &   5.770   &   245 &   6.603   &   246 \\
0.886   &   0.0202  &   1.595   &   3.02    &   2.444   &   38.6    &   3.279   &   104 &   4.122   &   7.28    &   4.937   &   98.7    &   5.779   &   235 &   6.613   &   245 \\
0.906   &   0.0287  &   1.609   &   2.12    &   2.454   &   34.5    &   3.289   &   110 &   4.132   &   6.81    &   4.947   &   121 &   5.789   &   228 &   6.623   &   243 \\
0.925   &   0.0384  &   1.619   &   2.17    &   2.464   &   32.3    &   3.299   &   116 &   4.142   &   6.61    &   4.957   &   129 &   5.799   &   223 &   6.633   &   246 \\
0.945   &   0.0532  &   1.629   &   2.23    &   2.473   &   28.6    &   3.309   &   107 &   4.152   &   6.37    &   4.967   &   124 &   5.809   &   219 &   6.643   &   250 \\
0.965   &   0.0801  &   1.639   &   2.29    &   2.483   &   26.2    &   3.319   &   103 &   4.162   &   6.20    &   4.976   &   110 &   5.819   &   218 &   6.653   &   252 \\
0.975   &   0.0893  &   1.649   &   2.38    &   2.493   &   24.7    &   3.329   &   93.4    &   4.172   &   6.19    &   4.986   &   95.8    &   5.829   &   219 &   6.662   &   254 \\
0.985   &   0.101   &   1.660   &   2.45    &   2.503   &   23.1    &   3.338   &   84.1    &   4.183   &   6.10    &   4.996   &   90.2    &   5.839   &   220 &   6.672   &   259 \\
0.994   &   0.116   &   1.670   &   2.75    &   2.513   &   21.7    &   3.348   &   79.7    &   4.193   &   6.03    &   5.006   &   92.4    &   5.850   &   224 &   6.682   &   258 \\
1.005   &       0.152   &       1.680   &       2.85    &       2.524   &       21.0    &       3.358   &       69.9    &       4.203   &       6.07    &       5.016   &       93.1    &       5.860   &       230     &       6.692   &       264     \\
1.016   &   0.164   &   1.690   &   2.95    &   2.534   &   20.5    &   3.368   &   68.9    &   4.213   &   6.17    &   5.026   &   95.6    &   5.870   &   236 &   6.702   &   268 \\
1.025   &   0.189   &   1.699   &   3.24    &   2.544   &   20.7    &   3.378   &   68.8    &   4.223   &   6.23    &   5.036   &   99.0    &   5.880   &   244 &   6.712   &   273 \\
1.030   &   0.209   &   1.709   &   3.34    &   2.554   &   21.1    &   3.388   &   76.4    &   4.232   &   6.43    &   5.046   &   103 &   5.889   &   265 &   6.722   &   279 \\
1.046   &       0.461   &       1.719   &       3.73    &       2.565   &       24.7    &       3.399   &       88.6    &       4.242   &       6.59    &       5.056   &       109     &       5.899   &       291     &       6.732   &       287     \\
1.050   &       0.850   &       1.730   &       4.11    &       2.576   &       38.3    &       3.409   &       85.6    &       4.252   &       6.86    &       5.066   &       114     &       5.909   &       346     &       6.752   &       307     \\
1.052   &       2.03    &       1.740   &       4.30    &       2.586   &       51.1    &       3.418   &       68.8    &       4.262   &       7.15    &       5.076   &       119     &       5.919   &       389     &       6.771   &       329     \\
1.053   &       4.53    &       1.749   &       4.37    &       2.595   &       46.2    &       3.427   &       48.0    &       4.272   &       7.70    &       5.086   &       124     &       5.929   &       389     &       6.791   &       353     \\
1.054   &       8.77    &       1.759   &       4.42    &       2.602   &       32.1    &       3.436   &       34.7    &       4.282   &       8.55    &       5.096   &       131     &       5.939   &       366     &       6.811   &       392     \\
1.055   &       4.54    &       1.769   &       4.81    &       2.611   &       24.9    &       3.446   &       25.7    &       4.292   &       9.70    &       5.106   &       137     &       5.939   &       401     &       6.831   &       438     \\
1.056   &       2.04    &       1.779   &       4.98    &       2.621   &       21.5    &       3.457   &       20.7    &       4.302   &       11.0    &       5.115   &       144     &       5.948   &       363     &       6.851   &       469     \\
1.058   &       0.880   &       1.789   &       5.26    &       2.632   &       20.4    &       3.466   &       17.1    &       4.313   &       13.7    &       5.125   &       150     &       5.958   &       326     &       6.871   &       468     \\
1.062   &   0.556   &   1.799   &   5.48    &   2.643   &   22.7    &   3.476   &   14.2    &   4.323   &   18.1    &   5.135   &   165 &   5.968   &   272 &   6.890   &   430 \\
1.078   &   0.495   &   1.808   &   5.66    &   2.657   &   51.8    &   3.486   &   12.2    &   4.333   &   26.7    &   5.145   &   169 &   5.978   &   254 &   6.910   &   416 \\
1.080   &   0.481   &   1.818   &   5.98    &   2.666   &   52.6    &   3.497   &   10.6    &   4.343   &   43.3    &   5.155   &   168 &   5.988   &   228 &   6.930   &   409 \\
1.087   &   0.466   &   1.828   &   6.42    &   2.673   &   39.6    &   3.507   &   9.18    &   4.353   &   71.7    &   5.165   &   167 &   5.997   &   209 &   6.950   &   437 \\
1.098   &   0.483   &   1.839   &   7.33    &   2.681   &   24.1    &   3.517   &   7.98    &   4.362   &   89.8    &   5.175   &   164 &   6.007   &   196 &   6.970   &   427 \\
1.112   &   0.485   &   1.849   &   7.87    &   2.690   &   21.1    &   3.527   &   7.09    &   4.371   &   76.0    &   5.185   &   161 &   6.018   &   196 &   6.989   &   437 \\
1.122   &   0.500   &   1.859   &   8.33    &   2.701   &   19.7    &   3.537   &   6.36    &   4.380   &   49.6    &   5.195   &   157 &   6.028   &   208 &   7.009   &   439 \\
1.132   &   0.561   &   1.869   &   8.77    &   2.712   &   19.6    &   3.547   &   5.80    &   4.389   &   31.7    &   5.205   &   154 &   6.038   &   214 &   7.029   &   447 \\
1.142   &   0.598   &   1.879   &   9.50    &   2.723   &   21.5    &   3.557   &   5.21    &   4.399   &   22.3    &   5.215   &   150 &   6.048   &   235 &   7.049   &   421 \\
1.152   &   0.614   &   1.889   &   10.1    &   2.735   &   35.5    &   3.567   &   4.71    &   4.410   &   16.9    &   5.224   &   146 &   6.058   &   296 &   7.069   &   398 \\
1.162   &   0.640   &   1.899   &   10.7    &   2.745   &   47.5    &   3.576   &   4.10    &   4.420   &   13.7    &   5.234   &   144 &   6.068   &   411 &   7.089   &   418 \\
1.172   &   0.664   &   1.909   &   11.4    &   2.754   &   41.9    &   3.586   &   3.60    &   4.430   &   12.1    &   5.244   &   142 &   6.079   &   564 &   7.098   &   457 \\
1.182   &   0.681   &   1.918   &   12.3    &   2.762   &   31.2    &   3.596   &   3.19    &   4.441   &   11.4    &   5.254   &   141 &   6.088   &   628 &   7.109   &   531 \\
1.192   &   0.740   &   1.928   &   13.7    &   2.772   &   45.9    &   3.606   &   2.89    &   4.451   &   10.1    &   5.264   &   150 &   6.098   &   565 &   7.109   &   507 \\
1.202   &   0.749   &   1.938   &   14.8    &   2.787   &   108 &   3.618   &   4.55    &   4.459   &   7.70    &   5.275   &   201 &   6.106   &   466 &   7.118   &   455 \\
1.212   &   0.784   &   1.948   &   15.8    &   2.792   &   115 &   3.630   &   10.2    &   4.469   &   6.59    &   5.284   &   228 &   6.116   &   399 &   7.128   &   410 \\
1.222   &   0.795   &   1.958   &   17.3    &   2.796   &   108 &   3.638   &   9.34    &   4.480   &   6.74    &   5.294   &   183 &   6.126   &   357 &   7.148   &   396 \\
1.232   &   0.796   &   1.968   &   18.8    &   2.801   &   60.1    &   3.645   &   5.85    &   4.490   &   7.05    &   5.303   &   156 &   6.136   &   312 &   7.168   &   382 \\
1.242   &   0.834   &   1.978   &   20.5    &   2.808   &   30.0    &   3.655   &   4.63    &   4.500   &   7.51    &   5.314   &   151 &   6.146   &   280 &   7.188   &   385 \\
1.251   &   0.868   &   1.988   &   22.6    &   2.818   &   22.6    &   3.666   &   4.31    &   4.510   &   7.99    &   5.324   &   152 &   6.157   &   268 &   7.207   &   373 \\
1.261   &   0.869   &   1.998   &   24.6    &   2.830   &   19.7    &   3.677   &   4.97    &   4.521   &   10.1    &   5.333   &   158 &   6.167   &   259 &   7.227   &   369 \\
1.271   &   0.866   &   2.009   &   28.0    &   2.841   &   18.0    &   3.691   &   19.1    &   4.531   &   10.9    &   5.343   &   168 &   6.177   &   263 &   7.247   &   374 \\
1.281   &   0.931   &   2.019   &   30.6    &   2.851   &   17.1    &   3.699   &   33.3    &   4.541   &   17.1    &   5.353   &   178 &   6.187   &   258 &   7.267   &   376 \\
1.291   &   0.887   &   2.028   &   33.6    &   2.862   &   16.7    &   3.706   &   19.4    &   4.549   &   55.1    &   5.363   &   193 &   6.197   &   266 &   7.287   &   381 \\
1.301   &   0.941   &   2.038   &   37.0    &   2.872   &   16.5    &   3.713   &   7.44    &   4.554   &   80.7    &   5.373   &   211 &   6.207   &   267 &   7.307   &   411 \\
1.303   &   0.918   &   2.048   &   40.3    &   2.882   &   16.5    &   3.722   &   4.16    &   4.558   &   89.6    &   5.383   &   225 &   6.216   &   271 &   7.326   &   438 \\
1.305   &   0.930   &   2.057   &   43.1    &   2.892   &   17.0    &   3.733   &   2.75    &   4.561   &   76.2    &   5.393   &   238 &   6.226   &   266 &   7.346   &   441 \\
1.307   &   0.950   &   2.067   &   44.2    &   2.902   &   17.0    &   3.744   &   2.31    &   4.564   &   57.7    &   5.403   &   243 &   6.236   &   265 &   7.366   &   454 \\
1.309   &   0.958   &   2.077   &   44.5    &   2.911   &   17.3    &   3.755   &   2.03    &   4.568   &   45.0    &   5.413   &   240 &   6.246   &   260 &   7.386   &   445 \\
1.311   &   0.990   &   2.087   &   48.5    &   2.921   &   17.6    &   3.765   &   1.72    &   4.578   &   37.1    &   5.423   &   237 &   6.256   &   267 &   7.406   &   456 \\
1.333   &   2.21    &   2.097   &   45.2    &   2.931   &   18.1    &   3.775   &   1.60    &   4.588   &   27.4    &   5.433   &   235 &   6.266   &   274 &   7.427   &   484 \\
1.334   &   4.26    &   2.106   &   41.2    &   2.941   &   18.7    &   3.786   &   1.53    &   4.598   &   21.6    &   5.443   &   240 &   6.276   &   284 &   7.445   &   502 \\
1.335   &   10.8    &   2.116   &   39.5    &   2.951   &   19.8    &   3.795   &   1.50    &   4.609   &   20.1    &   5.452   &   251 &   6.286   &   301 &   7.465   &   500 \\
1.336   &   48.2    &   2.126   &   38.2    &   2.962   &   21.1    &   3.805   &   1.47    &   4.619   &   20.2    &   5.462   &   254 &   6.296   &   314 &   7.485   &   509 \\
1.337   &   10.9    &   2.136   &   37.4    &   2.972   &   22.1    &   3.815   &   1.42    &   4.629   &   21.0    &   5.472   &   240 &   6.306   &   326 &   7.505   &   498 \\
1.338   &   4.27    &   2.146   &   37.5    &   2.982   &   24.4    &   3.825   &   1.39    &   4.639   &   22.5    &   5.482   &   202 &   6.316   &   349 &   7.525   &   498 \\
1.339   &   2.22    &   2.156   &   39.8    &   2.992   &   27.6    &   3.835   &   1.36    &   4.649   &   24.2    &   5.492   &   173 &   6.325   &   356 &   7.544   &   491 \\
1.365   &   1.14    &   2.167   &   42.8    &   3.002   &   31.9    &   3.845   &   1.39    &   4.659   &   25.9    &   5.502   &   151 &   6.335   &   362 &   7.564   &   450 \\
1.377   &   1.13    &   2.177   &   47.0    &   3.012   &   37.3    &   3.855   &   1.37    &   4.670   &   28.1    &   5.512   &   135 &   6.345   &   358 &   7.584   &   431 \\
1.389   &   1.12    &   2.187   &   53.0    &   3.021   &   41.6    &   3.865   &   1.43    &   4.679   &   29.3    &   5.522   &   123 &   6.355   &   344 &   7.605   &   430 \\
1.400   &   1.09    &   2.197   &   59.9    &   3.031   &   45.8    &   3.875   &   1.47    &   4.688   &   29.9    &   5.532   &   114 &   6.365   &   337 &   7.625   &   452 \\
1.410   &   1.11    &   2.207   &   66.4    &   3.041   &   47.6    &   3.885   &   1.57    &   4.698   &   30.3    &   5.542   &   107 &   6.375   &   322 &   7.645   &   499 \\
1.420   &   1.16    &   2.217   &   75.2    &   3.051   &   46.1    &   3.895   &   1.71    &   4.708   &   29.8    &   5.552   &   101 &   6.385   &   308 &   7.664   &   555 \\
1.431   &   1.21    &   2.227   &   75.3    &   3.060   &   43.8    &   3.905   &   1.84    &   4.718   &   30.0    &   5.561   &   96.6    &   6.395   &   296 &   7.684   &   613 \\
1.441   &   1.19    &   2.236   &   74.1    &   3.070   &   41.2    &   3.915   &   2.01    &   4.728   &   29.8    &   5.571   &   94.1    &   6.405   &   273 &   7.704   &   680 \\
1.450   &   1.22    &   2.245   &   73.3    &   3.080   &   38.8    &   3.925   &   2.28    &   4.738   &   29.6    &   5.581   &   92.0    &   6.415   &   261 &   7.724   &   691 \\
1.461   &   1.29    &   2.255   &   67.7    &   3.090   &   37.4    &   3.935   &   2.61    &   4.748   &   30.1    &   5.591   &   91.1    &   6.425   &   255 &   7.744   &   664 \\
1.471   &   1.29    &   2.265   &   66.2    &   3.100   &   37.3    &   3.945   &   2.95    &   4.758   &   30.6    &   5.601   &   91.3    &   6.435   &   246 &   7.763   &   638 \\
1.480   &   1.31    &   2.275   &   59.1    &   3.111   &   36.9    &   3.955   &   3.43    &   4.768   &   31.0    &   5.611   &   91.8    &   6.444   &   243 &   7.783   &   614 \\
1.490   &   1.31    &   2.285   &   56.0    &   3.121   &   37.0    &   3.965   &   4.07    &   4.778   &   31.3    &   5.621   &   94.0    &   6.454   &   244 &   7.803   &   592 \\
1.500   &   1.37    &   2.295   &   55.1    &   3.130   &   37.9    &   3.975   &   4.62    &   4.787   &   29.8    &   5.631   &   97.3    &   6.464   &   246 &   7.823   &   596 \\
1.510   &   1.42    &   2.305   &   53.8    &   3.140   &   39.2    &   3.985   &   5.14    &   4.797   &   29.0    &   5.641   &   100 &   6.474   &   245 &   7.843   &   595 \\
1.520   &   1.44    &   2.315   &   53.7    &   3.150   &   40.4    &   3.995   &   5.62    &   4.807   &   27.0    &   5.651   &   107 &   6.484   &   240 &   7.863   &   599 \\
1.530   &   1.57    &   2.325   &   52.9    &   3.160   &   42.3    &   4.005   &   6.20    &   4.817   &   25.5    &   5.662   &   115 &   6.494   &   241 &   7.882   &   622 \\
1.541   &   1.67    &   2.335   &   54.3    &   3.170   &   44.4    &   4.014   &   6.55    &   4.827   &   24.9    &   5.671   &   127 &   6.504   &   243 &   7.902   &   640 \\
1.550   &   1.62    &   2.346   &   56.5    &   3.180   &   47.2    &   4.024   &   7.05    &   4.837   &   24.8    &   5.681   &   142 &   6.514   &   244 &   7.922   &   583 \\
1.560   &   1.75    &   2.355   &   59.0    &   3.190   &   52.4    &   4.034   &   7.59    &   4.848   &   25.8    &   5.691   &   163 &   6.524   &   245 &   7.941   &   501 \\
1.585   &   2.90    &   2.365   &   63.3    &   3.200   &   56.9    &   4.044   &   8.83    &   4.858   &   27.2    &   5.701   &   182 &   6.534   &   244 &   7.962   &   458 \\
1.587   &       3.20    &       2.375   &       66.4    &       3.210   &       61.2    &       4.055   &       12.5    &       4.868   &       29.6    &       5.711   &       204     &       6.544   &       240     &               &              \\
\end{longtable*}
\end{center}
%\newpage


\begin{thebibliography}{99}
\bibitem{1}
G. Fiorentini, private communication (2004);
T. Araki {\it et al.}, Nature 436, 499 (2005).
\bibitem{2}
J. K. Bair, F. X. Haas, Phys. Rev. C {\bf 7}, 1356 (1973).
\bibitem{3}
H. W. Drotleff, A. Denker, H. Knee, M. Soine, G. Wolf, J. W. Hammer,
U. Greife, C. Rolfs, H. P. Trautvetter, Ap. J. {\bf 414}, 735 (1993).
\bibitem{4}
C. R. Brune, I. Licot, R. W. Kavanagh, Phys. Rev. C {\bf 48}, 3119 (1993).
\bibitem{5}
J. P. Biersack, J. F. Ziegler, \textit{Transport of Ions in Matter,
TRIM program version 2003} (IBM Research, New York, 1995)
\bibitem{6}
A. Denker, Ph.D thesis, Universit\"at Stuttgart, (1994) unpublished.
\bibitem{7}
C. E. Rolfs, W. Rodney, \textit{Cauldrons in the Cosmos}
(The University of Chicago Press, Chicago 1986)
%\bibitem{8} J. W. Hammer, private communication (2005).
% etc
\end{thebibliography}
\end{document}